# Gene Function Prediction with Gene Interaction Networks: A Context Graph Kernel Approach

Xin Li[1], *Student Member, IEEE*, Hsinchun Chen[2], *Fellow IEEE*, Jiexun Li[3], and Zhu Zhang[2]

[1] Department of Information Systems, City University of Hong Kong, 83 Tat Chee Avenue, Kowloon Tong, Hong Kong (xin.li@cityu.edu.hk)

[2] Department of Management Information Systems, University of Arizona, Tucson, AZ 85721, USA (hchen@eller.arizona.edu, zhuzhang@eller.arizona.edu)

[3] College of Information Science & Technology, Drexel University, Philadelphia, PA 19104, USA (jiexun.li@ischool.drexel.edu)

**Abstract**

Predicting gene functions is a challenge for biologists in the post-genomic era. Interactions among genes and their products compose networks that can be used to infer gene functions. Most previous studies adopt a linkage assumption, i.e., they assume that gene interactions indicate functional similarities between connected genes. In this study, we propose to use a gene's context graph, i.e., the gene interaction network associated with the focal gene, to infer its functions. In a kernel-based machine learning framework, we design a context graph kernel (CGK) to capture the information in context graphs. Our experimental study on a testbed of p53-related genes demonstrates the advantage of using indirect gene interactions and shows the empirical superiority of the proposed approach over linkage-assumption-based methods, such as GAIN and diffusion kernels.

**Index Terms**

Classification, kernel-based method, gene interaction network, gene function, gene pathway, system biology

## 1. Introduction

Developments in genome sequencing have led to the identification of a large number of genes. However, most of these genes' functions remain poorly known or unknown [1]. Annotating genes' functions has become a major challenge for biologists in the post-genomic era. Various computational techniques have been proposed to assist biological experiments in uncovering unknown gene functions.

In the early stages of computational modeling, individual genes' physical, chemical, and biological characteristics were the major features used for function prediction. Recent studies have used gene interaction information and obtained promising results [2]. However, in most of these studies gene interactions are considered to be indicators of functional similarities between connected genes, which restrict the power of prediction models. In addition, the topology of gene interaction networks has seen limited use.

In this paper, we propose to predict a gene's functions according to its context graph, which is defined as the gene interaction network composed of the genes interacting directly and indirectly with the focal gene. We propose a context graph kernel in a kernel-based machine learning framework that uses both gene features and structural characteristics of the context graph to infer the focal gene's functions.

The remainder of this paper is organized as follows. Section 2 reviews related studies on gene function prediction using gene interaction information. Section 3 introduces the proposed context graph kernel method. Section 4 describes our experiments and results. Section 5 summarizes our conclusions and future directions.

## 2. Literature Review

Gene functions can be predicted through annotating individual genes [3] or gene clusters [4]. At the individual gene level, gene features such as gene sequences [5], molecule structures [6], and gene co-expression patterns [7] have been used for annotation. Recent studies observed that gene interactions in biological pathways (including gene-gene interactions, gene-protein interactions, and protein-protein interactions) are also related to the functions of genes. The significant amount of gene interactions [8] found in previous research can become another important resource for gene function prediction.

We review previous interaction-based function prediction studies along three dimensions: assumptions, levels of interactions, and computational techniques.

**2.1 Assumptions**

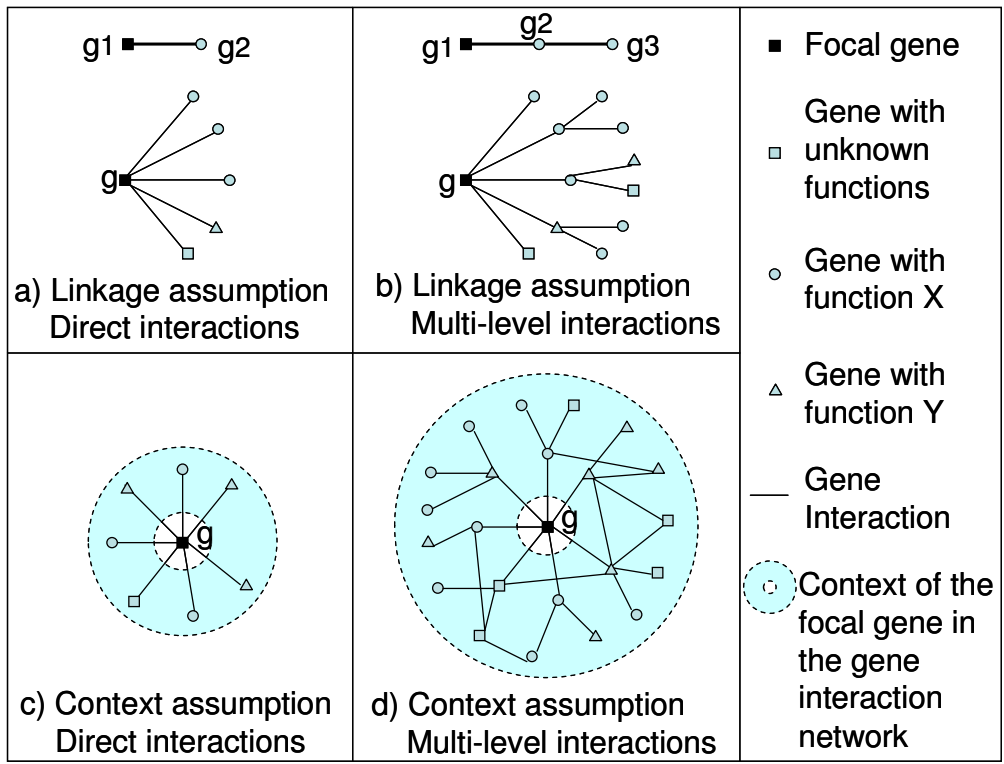

Fig. 1. Using gene interactions in gene function prediction. a) A linkage assumption assumes connected genes have similar functions. b) Indirect neighbors may have lower probability of sharing similar functions. c) A context assumption assumes that a gene's functions are correlated with the patterns of its context. d) When multiple levels of gene interactions are used, genes with similar context graphs may have similar functions.

Previous studies are typically built upon a *linkage assumption* or a *context assumption*. A linkage assumption considers gene interactions as an indicator of a functional similarity between connected genes. This assumption comes from the observations that immediate neighbor genes [9] and level-2 neighbor genes [10, 11] have a high probability of sharing functions. Based on this assumption, a focal gene's functions can be adopted from the majority of its neighbor's functions (Figures 1a and 1b).

The context assumption focuses on the relationship between a focal gene's functions and the pattern of its context, i.e., its direct/indirect neighbors' functions, sequences, or other characteristics. As illustrated in Figures 1c, gene g can be predicted to have a certain function based on the fact that it interacts with genes having function X and Y. However, it may not have function X or Y. For example, growth factor receptors (such as EGFR) usually bind to a growth factor (such as EGF) and activate certain proteins (e.g., SHC) and kinases (e.g., PI3K) in signaling pathways [12]. From a gene function prediction perspective, if a gene/gene product (such as PDGFR) is observed to be involved in a similar context of biological processes (binding to growth factors and activating SHC) may thus be predicted to be a growth factor receptor [13]. However, if the linkage assumption is taken, one may predict the gene to be a growth factor due to its interactions with other growth factor genes, which lead to a false prediction. In previous research, Schlitt et al. have considered using direct neighbors as the context and predicted similar functions for genes with similar neighbors [14].

**2.2 Levels of Interactions**

In previous gene function prediction research using gene interactions, both *direct interactions* between neighbor genes and *multiple levels of interactions* between indirect neighbor genes have been considered. Early studies based on direct interactions used the "guilt by association" rule under the linkage assumption to infer a focal gene's functions as the most frequent ones among its neighbors [15, 16].

Considering multiple levels of interactions is a natural extension. Under the linkage assumption, the indirect neighbors may have weaker influence on the focal gene's function prediction (Figure 1b). Features describing second-level neighbors have been explicitly extracted and used in gene function prediction [10, 11, 17]. Hishigaki et al. proposed searching multiple levels of neighbors for the most frequent functions to predict labels of the focal gene [18]. Furthermore, some studies extend the scope of neighbors to the entire gene interaction network to take advantage of the global topology of the network [8, 19, 20]. Multi-level interactions can also be considered under the context assumption, where the context graph extends beyond direct neighbors of the focal gene (Figure 1d) and is used in determining the focal gene's functions.

**2. 3 Computational Techniques**

Computational techniques for gene function prediction can be categorized into *heuristic approaches* and *machine learning approaches.* Heuristic approaches usually predefine rules to make predictions. For example, after defining the label propagation rule, function labels can be propagated through direct interactions [15] or the entire interaction network [20] to the genes with unknown functions. The predefined rules can also be used to design objective functions for optimization models. Based on the "guilt by association" rule, function prediction is formulated as minimizing inconsistent function assignments of connected genes in the network [8, 19, 21, 22]. Simulated annealing [19] and iterative local search methods [8, 21, 22] have been proposed to find solutions for such models.

Machine learning approaches build prediction models from training instances. In particular, kernel-based methods have been frequently used in gene function prediction, due to their ability to capture gene interaction network structures. Based on the context assumption, linear kernels [23] and graph overlap similarities [24] have been used to model focal genes' contexts. Based on the linkage assumption, diffusion kernels have been used to model genes' positional characteristics in gene networks [25, 26].

**2.4 Research Gaps**

Table 1 summarizes previous studies that use gene interactions in gene function prediction. We identify the following research gaps in previous research:

- Most studies tackled the gene function prediction problem under a linkage assumption. The use of context assumption is limited.

- Although both direct interactions and multi-level interactions have been used under the linkage assumption, the context-based studies used only direct interactions. The effect of indirect interactions under the context assumption still remains to be investigated.

- Heuristic approaches have been the major technique adopted. Several machine-learning-based studies used features of individual genes [5-7] without considering gene interactions. However, less attention has been paid to leveraging graph structures in statistical learning for gene function prediction.

Table 1. A summary of previous studies

| Previous studies | Alg. Type* | Descriptions |
|---|---|---|
| Mayer et al. 2000 [15] | L/D/H | Guilt by association (majority voting) |
| Schwikowski et al. 2000 [9] | L/D/H | Guilt by association |
| Hishigaki et al. 2001 [18] | L/M/H | Multi-level neighbor majority voting |
| Schlitt et al. 2003 [14] | C/D/H | Similarity of neighbors |
| Vazquez et al. 2003 [19] | L/M/H | Minimize un-matching gene pairs |
| Karaoz, U. et al. 2004 [8] | L/M/H | Minimize un-matching gene pairs |
| Lanckriet et al. 2004 (a) [23] | L/M/ML | Diffusion kernel (classification by gene positions in the network) |
| Lanckriet et al. 2004 (b) [23] | C/D/ML | Linear kernel |
| Tsuda et al. 2004 [26] | L/M/ML | Locally constraint diffusion kernel |
| Yamanishi et al. 2004 [25] | L/M/ML | Diffusion kernel |
| Nabieva et al. 2005 [20] | L/M/H | Label propagation |
| Massjouni et al. 2006 [21] | L/M/H | Minimize un-matching gene pairs |
| Chua et al. 2006 [10] | L/M/H | Weighted neighbor function label counting |
| Murali et al. 2006 [22] | L/M/H | Minimize un-matching gene pairs |
| Xu et al. 2006 [17] | B/M/ML | KNN with neighbor-related features |
| Chua et al. 2007 [11] | L/M/H | Weighted neighbor function label counting |
| Li et al. 2007 [16] | L/D/H | Guilt by association |
| Zhao et al. 2008 [24] | C/D/H | Similarity of neighbors |

Note*: The three symbols separated by slashes represent the assumption, level of interactions, and computational technique respectively. The symbols' meanings: L – linkage assumption; C – context assumption; B – both linkage assumption and context assumption; D – direct interactions; M – multi-level interactions; H – heuristic approach; ML – machine learning approach.

## 3. Research Design

To bridge the aforementioned research gaps, our study aims at predicting a gene's functions based on its context in a gene interaction network. We also inspect the effect of using multiple levels of (indirect) interactions in function prediction. We choose a kernel-based machine learning approach in this research due to its documented good performance and ability to handle structural data [27].

### 3.1 A Kernel-based Approach

We formulate gene function prediction as a classification problem, i.e., categorizing genes into classes of gene functions. We first extract biological interactions from public databases to build interaction networks. In this research, we study only gene-gene interactions. Thus, for the databases for interactions between proteins or other gene products, we map all gene products to genes and consolidate the interactions to gene-level. While our proposed framework can also be applied to infer the functions of proteins in protein-protein interactions, we leave that to future research. We then annotate genes with their known function labels. The genes with known functions are used as training instances to build classification models. Specifically, we build a binary classifier for each function label to test whether a gene has this function. Predictions on the testing instances by the classifiers can be validated against existing knowledge by domain experts or through further experiments.

In a kernel-based machine learning framework, we need to specify a kernel function and a kernel machine. The kernel function defines a similarity measure between data instances (i.e., genes in our research), while the kernel machine is in charge of building the model to classify the data instances. The performance of kernel-based methods is highly dependent on the design of kernel functions [28]. Thus, the main focus and contribution of the paper are to design a kernel function that can better capture structural patterns in gene interaction networks for the gene function prediction task. For the kernel machine, we choose the Support Vector Machines (SVM) [29] algorithm due to its reported competitive performance [30, 31].

### 3.2 A Context Graph Kernel

#### 3.2.1 Kernel Design

Recognizing the limitations of previous research, we adopt a context assumption and use multiple levels of gene interactions for gene function prediction. As shown in Figure 1d, we represent each gene's context as a graph. A context graph centers on the focal gene and includes its direct and indirect neighbors. According to the context assumption, genes with similar context graphs may share similar functions. Therefore we design a context graph kernel (CGK) to compute the similarity between context graphs.

Various graph kernels have been developed in previous data mining studies. For example, graph kernels were used to classify proteins based on the similarity of their molecular structures [6, 32, 33]. These graph kernels belong to the family of convolution kernels [34] and compute the graph similarity by accumulating the similarity scores of random walk paths on the graphs in a pair-wise manner.

Our proposed CGK also relies on the comparisons of random walk paths in the graphs. However, unlike traditional graph kernels that utilize all random walk paths in the graph, CGK considers only the random walk paths that start from the focal gene. This design is due to our objective of predicting the focal gene's functions. Figure 2 shows the context graph of gene $g_0$ and random walk paths starting from this focal gene. These paths represent the gene pathways related to the focal gene and thus may potentially indicate its functions. We calculate the similarity between two context graphs as the sum of pair-wise similarities of these random walk paths. Each path's contribution is weighted according to its probability of traversal. Since longer random walk paths have a relatively lower probability of being traversed, the genes that are far from the focal gene have less impact on the focal gene's function.

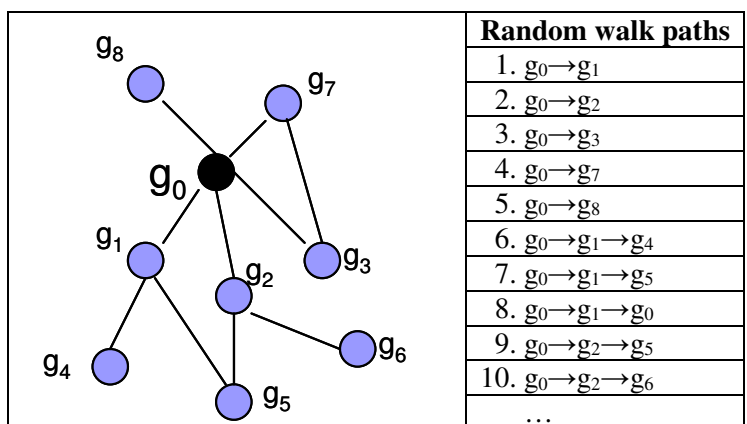

Fig. 2. Random walk paths on a context graph

The following procedure summarizes our kernel design:

(I) In the gene interaction network $G$ of a genome with $n$ genes $\{g_1, g_2, \ldots g_n\}$, we represent gene $g_x$'s context graph as $G_x$. All random walks in $G_x$ start at $g_x$. At each gene (node) $g_i$, a random walk has a probability of $p_s(g_i)$ to stop and a probability of $p_t(g_j|g_i)$ to jump to one of $g_i$'s neighbors, $g_j$. Thus, a random walk path of length *l*, $h=(g_{<h,0>} \rightarrow g_{<h,1>} \rightarrow ... \rightarrow g_{<h,l>})$, has the probability of traversal:

$$P(h|G_x) = p_t(g_{<h,1>}|g_{<h,0>})p_t(g_{<h,2>}|g_{<h,1>}) \cdots p_t(g_{<h,l>}|g_{<h,l-1>})p_s(g_l) \qquad (1)$$

where $g_{<h,i>}$ indicates the *i*-th gene on the path *h*.

(II) After enumerating all random walk paths, the similarity between two context graphs, $K(G_x,G_y)$ is defined as the sum of the similarity scores of random walk paths weighted by the paths' probability of traversal:

$$K(G_x,G_y) = \sum_{h_i \in H(G_x)} \sum_{h_j \in H(Gy)} K_h(h_i,h_j)P(h_i|G_x)P(h_j|G_y) \qquad (2)$$

where $K_h(h_i,h_j)$ is the similarity score between two random walk paths $h_i$ and $h_j$, and $H(G_x)$ and $H(G_y)$ denote the sets of random walk paths in the two context graphs.

(III) The similarity between random walk paths is computed by multiplying the similarity of the corresponding nodes along the two paths. We define $K_h(h_i,h_j)$ as:

$$K_h(h_i,h_j) = \begin{cases} \prod_k K_g(g_{<h_i,k>}, g_{<h_j,k>}), & if \ |h_i|=|h_j| \\ 0, & otherwise \end{cases} \qquad (3)$$

where $|h_i|$ and $|h_j|$ are the lengths of the two paths $h_i$ and $h_j$, and $K_g(g_{<h_i,k>}, g_{<h_j,k>})$ is a similarity function defined on nodes.

(IV) The node similarity function $K_g(g_{<h_i,k>}, g_{<h_j,k>})$ captures the node-level information of individual genes, potentially available from experiments, literature, gene sequences, and ontologies [5, 18-20, 22]. It can be defined in various ways. Since finding the best node similarity function is not our main research focus, we adopt a simple function: . If two genes share at least one common neighbor gene, the similarity between the two genes is set as 1. Otherwise, the similarity is set as 0.

(V) In practice, the CGK kernel can be normalized for better classification performance.

$$K'(G_x,G_y) = K(G_x,G_y) / \sqrt{K(G_x,G_x)K(G_y,G_y)} \qquad (4)$$

**3.2.2 Computing the Context Graph Kernel in a Matrix Form**

To compute the CGK by enumerating all random walk paths is computationally expensive. We introduce an efficient method for computing the CGK based on the matrix form of the kernel.

In the gene interaction network $G$ of a genome with $n$ genes $\{g_1, g_2, \ldots g_n\}$, we use two n*n matrices $M=\{M_{i,j}\}=\{p_t(g_j|g_i)\}$ and $Q=\{Q_{i,j}\}=\{p_s(g_j)\}$ to encode each node's transition probability and stopping probability in the graph, respectively. The context graph kernel matrix of the entire genome $\tilde{K} = \{\tilde{K}_{i,j}\} = \{K(G_i, G_j)\}$ can be represented as the summation of a series of matrices (Proposition 1 in Appendix):

$$\tilde{K} = \sum K_i \quad (i=1,2,\ldots,\infty) \qquad (5)$$

where $K_1=(M*Q)K_0(M*Q)^T$, $K_{i+1}=M(K_0*K_i)M^T$ $(i=1,2,\ldots,\infty)$, and $K_0 = \{K_{0_{i,j}}\} = \{K_g(g_i, g_j)\}$ is the kernel matrix of node information. The operation * is the Hadamard product (i.e., entrywise product) where $A*B=\{a_{i,j} \cdot b_{i,j}\}$.

Each matrix $K_i$ covers the random walk paths of length $= i$ in the context graphs. It can be proved that $K_1+K_2+\ldots+K_r$ converges when $r$ approaches $+\infty$ if the stopping probability is uniform or larger than 0.5 on all nodes (Proposition 2 in Appendix). Therefore, given a maximum length of random walk paths, $r$, we can use $K_1+K_2+\ldots+K_r$ to approximate the kernel $\tilde{K}$. With the decomposed form of $K_i$, the kernel $\tilde{K}$ can be computed by simple matrix operations with a time complexity in $O(rn^3)$, where $n$ is the number of nodes in the network.

The matrix formulation also facilitates the investigation of the effect of indirect interactions on gene function prediction. Since all random walks start from the focal genes, each $K_i$ covers the genes that are $i$ step(s) away from the focal gene. While $\sum K_i$ will converge as $r$ increases, specifying a different $r$ restricts the CGK to a limited number of indirect interactions and may yield a different prediction performance. This may help us understand the effect of indirect interactions in gene function prediction.

## 4. Experimental Study

### 4.1 Dataset

**4.1.1 Human Genome Gene Interaction Network**

In this study we use the collection of gene interactions from the BioGRID database [35] to construct a gene interaction network of human genomes. BioGRID is a free and well-known database with protein/gene interactions manually curated from Medline literature. We extract 38,225 relations related to *Homo sapiens* genes from BioGRID (version 2.026). By mapping proteins to genes and consolidating duplicate relations, we construct a gene interaction network with 19,623 non-directional relations among 7,167 genes.

**4.1.2 Gene Function Labels**

Following previous studies and domain experts' suggestions, we use terms from Gene Ontology's "biological process" hierarchy (downloaded in 2009) [36] as gene function labels. In GO, each term is associated with an evidence code indicating how the annotation to the term is supported (http://www.geneontology.org/GO.evidence.shtml). For the evaluation of gene function prediction methods, we used only the terms based on biological experimental evidences and computational analysis evidences and excluded the terms without solid evidence (i.e., NAS, IC, ND, IEA, and NR) in our experiments. After pre-processing, the "biological process" hierarchy has 8 second-level terms and 125 third-level terms (first level is "biological process"). We use the third-level terms as class labels in our study so as to have both enough classification granularity and sufficient training/testing data instances for each class. The genes whose functions are not documented are annotated as "unknown."

**4.1.3 A p53-related Testbed**

The tumor suppressor gene, p53, plays a central role in the regulation of apoptosis and cell cycle arrest in cancer development. P53-related genes have attracted much attention and their functions are well-studied compared to other human genes. Therefore, we choose p53-related genes as our research testbed. In our previous research, we identified 2,045 p53-related genes from the Medline abstracts with a Natural Language Processing tool [37, 38]. After eliminating the genes without known functions, we obtain 1,566 genes within 35 third-level "biological process" functions. Nine of the classes with more than 50 instances are used in our experiments for evaluation (Table 1 in Appendix). The dataset contains 819 gene in total. It should be noted that although the training/testing instances only include these p53-related genes, the gene interaction network-based approaches in our experiments take advantage of all (known) genes and gene interactions in the human genome to build gene function prediction models.

**4.2 Evaluation Methodology**

In the implementation of CGK, we specify a uniform stopping probability $p_s(g_i)=1-\lambda$ $(0<\lambda<1)$ to generate random walks on the gene interaction network $G$. We also assume equal probability of jumping from one node to any of its neighbors, i.e., $p_t(g_j|g_i)=\lambda/d(g_i)$, where $d(g_i)$ is the number of $g_i$'s interacting genes. After the calculation and normalization of the CGK kernel, we use a popular SVM package, libSVM [39], to build classifiers.

We design the following two sets of experiments:

- *Experiment I* is to examine the effect of indirect interactions in gene function prediction. As described in 3.2.2, $r$ controls the levels of indirect interactions that can be used in the kernel computation. In addition, the stopping probability $p_s(g_i)=1-\lambda$ also controls the effect of indirect interactions in that longer random walks have smaller probabilities to be utilized in the kernel. Thus, we compare the performances of CGK using different $r$ and $\lambda$ settings in this experiment.

- *Experiment II* is to compare our proposed CGK-based method with other state-of-the-art methods, specifically four baseline methods from previous studies. In this experiment, we calculated the CGK kernel matrix till convergence ($r=6$ in our experiments).

To build a model with the CGK kernel, we need to specify the parameters of the SVM algorithm and $\lambda$ ($r$ does not need to be selected since one can calculate the kernel till converge). In the experiments, we use 50% of the data for parameter selection and the other 50% for performance testing. Thus, only 50% genes' functions are assumed known in each stage. Through a 5-fold cross validation, we set $\lambda$ to 0.9 (from 0.1 to 0.9), where the parameters for SVM are selected accordingly using the grid search tool provided by libSVM. With these parameters, the performance of the model on the testing dataset is measured also using a 5-fold cross validation. In the process of cross validation, the functions of the genes to be classified are considered as "unknown" in the gene interaction network.

**4.3 Evaluation Metrics**

We evaluate the performance of the classification models using precision, recall, and F-measure, which are common evaluation metrics in gene function prediction studies [3, 8, 22]. Since one gene may have more than one function and one function may be associated with more than one gene, we calculate the three measures at both the instance level (i.e., gene level) and class level (i.e., function level). We also inspect instance-level performance with respect to the number of interacting genes to better understand the algorithms' characteristics [18].

Instance-level precision $P_i$, recall $R_i$, and F-measure $F_i$ are defined as:

$$P_i = \frac{\text{correctly predicted functions of a gene}}{\text{all predicted functions of a gene}} \quad (6)$$

$$R_i = \frac{\text{correctly predicted functions of a gene}}{\text{all (known) functions of a gene}} \quad (7)$$

$$F_i = 2 \times P_i \times R_i / (P_i + R_i) \quad (8)$$

Class-level precision $P_c$, recall $R_c$, and F-measure $F_c$ are defined as:

$$P_c = \frac{\text{correctly predicted genes of a class}}{\text{all predicted genes of a class}} \quad (9)$$

$$R_c = \frac{\text{correctly predicted genes of a class}}{\text{all (known) genes of a class}} \quad (10)$$

$$F_c = 2 \times P_c \times R_c / (P_c + R_c) \quad (11)$$

**4.4 Experimental Results and Discussion**

**4.4.1 Experiment I: Effect of Indirect Interactions**

Experiment I examines the effect of indirect interactions for gene function prediction. Figure 3 shows the instance-level and class-level performances using different values of $r$ and $\lambda$. When only direct interactions are considered ($r$=1), the normalized CGK kernels are identical for different $\lambda$'s, which yield average F-measure scores of 25.5% at the instance level and 30.3% at the class level. When $\lambda$ is larger than 0.5, taking into account of an additional level of indirect interactions (*r*=2) leads to a significant increases in F-measure scores. When more levels of indirect interactions are included, the F-measure curves stabilize, which indicates the convergence of the kernel computation.

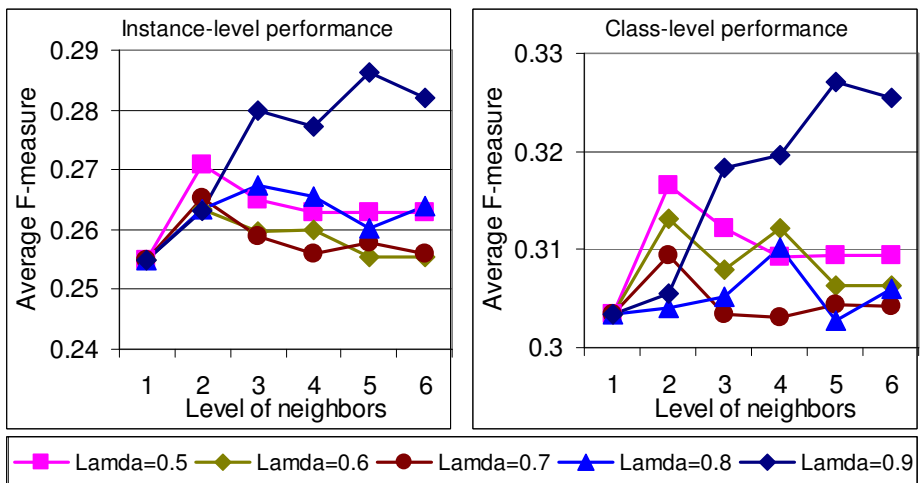

Fig. 3. Performance of CGK using different levels of interactions

It should be noted that a larger *r* indicates more indirect interactions used in our experiments. A larger λ indicates indirect interactions playing a more important role in the kernel. The experimental results demonstrate that incorporating information of indirect interactions may improve the gene function prediction performance. Our experiments suggest that computing CGKs for two or three levels (*r* = 2 or 3) gives a good approximation of the CGK, which can be considered in practical applications.

**4.4.2 Experiment II: CGK vs. Other Methods**

**4.4.2.1 Benchmark Algorithms**

Based on the three dimensions in our literature review (i.e., assumption, interactions, and techniques), in Experiment II we compare our proposed context graph kernel method against four baseline methods from previous studies:

  a) Linear kernel: A linear kernel (LK) uses direct interactions under a context assumption [23]. For gene function prediction, the genes directly connected to the focal gene are used as features to represent the gene. The inner product of the feature vectors is used to calculate genes' similarities. Genes with higher similarities, i.e., genes that share a larger number of neighbors, are predicted to have similar functions.

b) Diffusion kernel: A diffusion kernel (DK) uses multi-level interactions under a linkage assumption. DK uses genes' relative positions in the gene interaction network to predict their functions. Two genes that have more and shorter paths between them are more likely to be predicted to share the function labels [23, 25]. In addition to using the traditional diffusion kernel, we also adopt a locally constrained diffusion kernel (LDK) as proposed by Tsuda et al. due to its reported superior performance [26].

c) Gene Annotation using Interaction Network (GAIN): GAIN is a heuristic method using gene interaction networks under a linkage assumption, as reported in Karaoz et al. [8] and Murali et al. [22]. It optimizes gene function assignment by minimizing the inconsistency among connected genes in the network.

d) Neighbor majority voting method (MV): Guided by the "guilt by association" rule [15], we construct a simple neighbor majority voting classifier. The most dominant function label in the directly connected neighbors is predicted as the focal gene's function.

In this research, the implementation of LDK and GAIN were provided by their authors, while the others were implemented by us.

### 4.4.2.2 Instance-level Performance

Table 2. Instance-level classification performance

| *Instance-level performance* | *Average precision* | *Average recall* | *Average F-measure* |
|---|---|---|---|
| Context Graph Kernel (CGK) | **29.07%** | **33.51%** | **28.04%** |
| Linear Kernel (LK) | 21.54% | 18.37% | 18.40% |
| Diffusion Kernel (DK) | 23.48% | 21.69% | 20.80% |
| Locally Constrained Diffusion Kernel (LDK) | 20.90% | 18.97% | 18.31% |
| Gene Annotation using Interaction Network (GAIN) | **27.41%** | 23.70% | 23.82% |
| Majority Voting (MV) | 6.02% | 4.37% | 4.86% |

\* Within the same measure, the bold numbers do not have a significant difference from the largest one at the 95% confidence level.

Table 2 shows the instance-level prediction performances for different methods. Our proposed context graph kernel achieves the highest average precision, recall, and F-measure, which are significantly better than most other methods with a p-value < 0.05 in pair-wise *t*-tests. Its precision is not significantly different from GAIN (p-value≈0.26). Compared to all learning-based methods, the context graph kernel outperforms by about 10% on all three measures. In these experiments, the majority voting algorithm has significantly worse performance than the other algorithms, which is consistent with previous research [19].

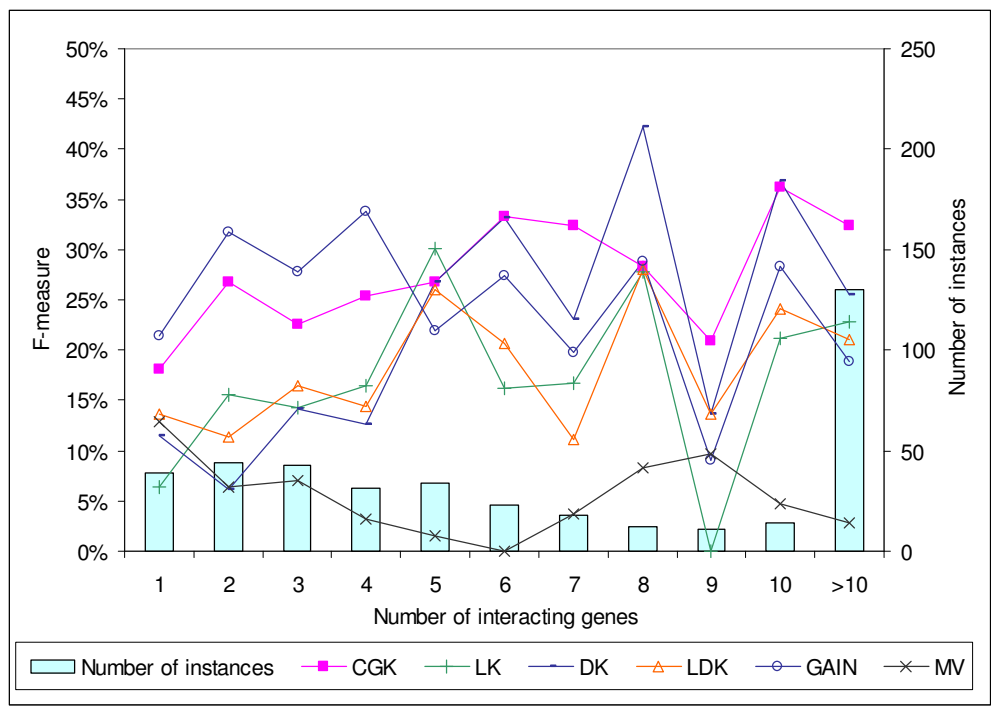

Fig. 4. Instance-level F-measure for genes with different interacting genes

Figure 4 (and Figure 1 in Appendix) shows different classifiers' performances for genes with different numbers of interacting genes (or in graph theoretical terms, nodes with different degrees). The number of genes in each group is shown by column bars. In general, there is a positive correlation between classification performance and the number of interacting genes, except for the majority voting algorithm and GAIN. Most algorithms were able to capture the information provided by a larger number of interactions for a more accurate prediction. Figure 4 shows that the CGK performance is always among the best of all, which shows its ability of classifying different types of genes.

**4.4.2.3 Class-level Performance**

Table 3. Class-level prediction performance

| GO term | | 42330 | 51649 | 44419 | 8283 | 6928 | 42221 | 43170 | 48869 | 48856 |
|---|---|---|---|---|---|---|---|---|---|---|
| # of genes | | 24 | 36 | 41 | 46 | 49 | 77 | 111 | 116 | 136 |
| CGK | $P_C$ | 80.95% | 12.50% | 10.53% | **11.32%** | 26.42% | 28.33% | 49.14% | 37.38% | 36.96% |
|  | $R_C$ | **70.83%** | 11.11% | 9.76% | **13.04%** | 28.57% | 22.08% | 51.35% | **34.48%** | 37.50% |
|  | $F_C$ | 75.56% | 11.76% | 10.13% | **12.12%** | 27.45% | 24.82% | 50.22% | **35.87%** | 37.23% |
| LK | $P_C$ | **100.00%** | 25.00% | **50.00%** | 8.33% | 25.00% | 12.00% | 57.63% | 34.62% | 42.27% |
|  | $R_C$ | 54.17% | 2.78% | 7.32% | 2.17% | 6.12% | 3.90% | 30.63% | 15.52% | 30.15% |
|  | $F_C$ | 70.27% | 5.00% | 12.77% | 3.45% | 9.84% | 5.88% | 40.00% | 21.43% | 35.19% |
| DK | $P_C$ | **100.00%** | 37.50% | 33.33% | 7.14% | **31.82%** | 27.27% | **66.67%** | 40.54% | 47.19% |
|  | $R_C$ | 54.17% | 8.33% | **12.20%** | 2.17% | 14.29% | 11.69% | 36.04% | 25.86% | 30.88% |
|  | $F_C$ | 70.27% | 13.64% | **17.86%** | 3.33% | 19.72% | 16.36% | 46.78% | 31.58% | **37.33%** |
| LDK | $P_C$ | **100.00%** | 0.00% | 16.67% | 7.14% | 28.57% | 25.93% | 62.07% | 47.17% | 50.00% |
|  | $R_C$ | 62.50% | 0.00% | 4.88% | 2.17% | 12.24% | 9.09% | 32.43% | 21.55% | 27.21% |
|  | $F_C$ | 76.92% | 0.00% | 7.55% | 3.33% | 17.14% | 13.46% | 42.60% | 29.59% | 35.24% |
| GAIN | $P_C$ | **100.00%** | **58.33%** | 22.22% | 0.00% | 16.67% | **35.29%** | 65.12% | **47.22%** | 56.25% |
|  | $R_C$ | **70.83%** | 19.44% | 4.88% | 0.00% | 4.08% | 7.79% | 50.45% | 14.66% | 19.85% |
|  | $F_C$ | 82.93% | **29.17%** | 8.00% | 0.00% | 6.56% | 12.77% | **56.85%** | 22.37% | 29.35% |
| MV | $P_C$ | **100.00%** | 11.11% | 11.76% | 0.00% | 28.57% | 22.22% | 42.86% | 36.36% | **60.00%** |
|  | $R_C$ | 16.67% | 2.78% | 4.88% | 0.00% | 4.08% | 2.60% | 5.41% | 3.45% | 2.21% |
|  | $F_C$ | 28.57% | 4.44% | 6.90% | 0.00% | 7.14% | 4.65% | 9.60% | 6.30% | 4.26% |

Table 3 shows the class-level performance of different classifiers. For most classes, the context graph kernel achieves the highest recall, while GAIN and the two diffusion kernels (DK and LDK) achieve the highest precision. All of these methods utilize the structure of the gene interaction network. Their performance differences show the different prediction power of the context assumption and the linkage assumption. In general, we observe a positive correlation between the class-level performance measures and the number of instances in the classes. For the classes with a larger number of instances, the F-measure of CGK is usually among the best. We identify that the class "GO:42221 (response to chemical stimulus)" is difficult to classify (compared with classes with a similar number of instances) for most of the algorithms. Function prediction methods based on gene interactions do not perform well on this class, probably because it is less related to gene interactions. Other types of information may need to be considered in this case.

## 5. Conclusions and Future Directions

In this research, we propose a context graph kernel for gene function prediction. This approach is based on a context assumption; it leverages multiple levels of interactions in gene interaction networks. Compared to other state-of-the-art methods that often use linkage assumptions and/or direct interactions, our proposed approach is highly competitive and achieves the highest F-measure. In addition, we find that our proposed approach works better on genes with a larger number of interacting genes and classes with a larger number of genes.

The context graph kernel is capable of incorporating different types of node information in the gene function prediction task. In the future, we will extend the current research to include other biological data such as node information. We also plan to study the gene function prediction problem in multi-genome gene interaction networks.

## Acknowledgment

This work was supported by: NIH/NLM, 1 R33 LM07299-01, "Genescene: a Toolkit for Gene Pathway Analysis." The authors would like to thank Dr. T. M. Murali (https://bioinformatics.cs.vt.edu/~murali/software/gain/) and Dr. K. Tsuda (http://kyb.tuebingen.mpg.de/bs/people/tsuda/ismb04.html) for making their algorithm source codes available to the research community. Their algorithms are used as baseline methods in our study.

## References

[1] A. J. Enright, V. Kunin, and C. A. Ouzounis, "Protein families and TRIBES in genome sequence space," *Nucleic Acids Research,* vol. 31, pp. 4632-4638, AUG 1 2003.

[2] P. Z. Hu, G. Bader, D. A. Wigle, and A. Emili, "Computational prediction of cancer-gene function," *Nature Reviews Cancer,* vol. 7, pp. 23-34, JAN 2007.

[3] R. Sharan, I. Ulitsky, and R. Shamir, "Network-based prediction of protein function," *Molecular Systems Biology,* vol. 3, MAR 2007.

[4] M. A. Huynen, B. Snel, C. von Mering, and P. Bork, "Function prediction and protein networks," *Current Opinion in Cell Biology,* vol. 15, pp. 191-198, APR 2003.

[5] L. J. Jensen, R. Gupta, H. H. Staerfeldt, and S. Brunak, "Prediction of human protein function according to Gene Ontology categories," *Bioinformatics,* vol. 19, pp. 635-642, MAR 22 2003.

[6] K. M. Borgwardt, C. S. Ong, S. Schonauer, S. V. N. Vishwanathan, A. J. Smola, and H. P. Kriegel, "Protein function prediction via graph kernels," *Bioinformatics,* vol. 21, pp. I47-I56, JUN 2005.

[7] P. Pavlidis, J. Weston, J. S. Cai, and W. S. Noble, "Learning gene functional classifications from multiple data types," *Journal of Computational Biology,* vol. 9, pp. 401-411, 2002.

[8] U. Karaoz, T. M. Murali, S. Letovsky, Y. Zheng, C. M. Ding, C. R. Cantor, and S. Kasif, "Whole-genome annotation by using evidence integration in functional-linkage networks," *Proceedings of the National Academy of Sciences of the United States of America,* vol. 101, pp. 2888-2893, MAR 2 2004.

[9] B. Schwikowski, P. Uetz, and S. Fields, "A network of protein-protein interactions in yeast," *Nature Biotechnology,* vol. 18, pp. 1257-1261, DEC 2000.

[10] H. N. Chua, W. K. Sung, and L. Wong, "Exploiting indirect neighbours and topological weight to predict protein function from protein-protein interactions," *Bioinformatics,* vol. 22, pp. 1623-1630, JUL 1 2006.

[11] H. N. Chua, W. K. Sung, and L. Wong, "Using indirect protein interactions for the prediction of Gene Ontology functions," *Bmc Bioinformatics,* vol. 8, 2007.

[12] C. J. Wu, Z. J. Chen, A. Ullrich, M. I. Greene, and D. M. O'Rourke, "Inhibition of EGFR-mediated phosphoinositide-3-OH kinase (PI3-K) signaling and glioblastoma phenotype by Signal-Regulatory Proteins (SIRPs)," *Oncogene,* vol. 19, pp. 3999-4010, AUG 17 2000.

[13]     R. A. Klinghoffer, B. Duckworth, M. Valius, L. Cantley, and A. Kazlauskas, "Platelet-derived growth factor-dependent activation of phosphatidylinositol 3-kinase is regulated by receptor binding of SH2-domain-containing proteins which influence Ras activity," *Molecular and Cellular Biology,* vol. 16, pp. 5905-5914, OCT 1996.

[14]     T. Schlitt, K. Palin, J. Rung, S. Dietmann, M. Lappe, E. Ukkonen, and A. Brama, "From gene networks to gene function," *Genome Research,* vol. 13, pp. 2568-2576, DEC 2003.

[15]     M. L. Mayer and P. Hieter, "Protein networks - built by association," *Nature Biotechnology,* vol. 18, pp. 1242-1243, DEC 2000.

[16]     Y. H. Li, Z. Guo, W. C. Ma, D. Yang, D. Wang, M. Zhang, J. Zhu, G. C. Zhong, Y. J. Li, C. Yao, and J. Wang, "Finding finer functions for partially characterized proteins by protein-protein interaction networks," *Chinese Science Bulletin,* vol. 52, pp. 3363-3370, DEC 2007.

[17]     J. Z. Xu and Y. J. Li, "Discovering disease-genes by topological features in human protein-protein interaction network," *Bioinformatics,* vol. 22, pp. 2800-2805, NOV 15 2006.

[18]     H. Hishigaki, K. Nakai, T. Ono, A. Tanigami, and T. Takagi, "Assessment of prediction accuracy of protein function from protein-protein interaction data," *Yeast,* vol. 18, pp. 523-531, APR 2001.

[19]     A. Vazquez, A. Flammini, A. Maritan, and A. Vespignani, "Global protein function prediction from protein-protein interaction networks," *Nature Biotechnology,* vol. 21, pp. 697-700, JUN 2003.

[20]     E. Nabieva, K. Jim, A. Agarwal, B. Chazelle, and M. Singh, "Whole-proteome prediction of protein function via graph-theoretic analysis of interaction maps," *Bioinformatics,* vol. 21, pp. I302-I310, JUN 2005.

[21]     N. Massjouni, C. G. Rivera, and T. M. Murali, "VIRGO: computational prediction of gene functions," *Nucleic Acids Research,* vol. 34, pp. W340-W344, JUL 1 2006.

[22]     T. M. Murali, C. J. Wu, and S. Kasif, "The art of gene function prediction," *Nature Biotechnology,* vol. 24, pp. 1474-1475, DEC 2006.

[23]     G. R. G. Lanckriet, T. De Bie, N. Cristianini, M. I. Jordan, and W. S. Noble, "A statistical framework for genomic data fusion," *Bioinformatics,* vol. 20, pp. 2626-2635, NOV 1 2004.

[24]    X. M. Zhao, Y. Wang, L. N. Chen, and K. Aihara, "Gene function prediction using labeled and unlabeled data," *Bmc Bioinformatics,* vol. 9, pp. -, JAN 28 2008.

[25]    Y. Yamanishi, J.-P. Vert, and M. Kanehisa, "Protein network inference from multiple genomic data: a supervised approach," *Bioinformatics,* vol. 20, pp. i363-i370, 2004.

[26]    K. Tsuda and W. S. Noble, "Learning kernels from biological networks by maximizing entropy," *Bioinformatics,* vol. 20, pp. i326-i333, 2004.

[27]    T. Gartner, "A survey of kernels for structured data," *ACM SIGKDD Explorations,* vol. 5, pp. 49-58, 2003.

[28]    Y. Tan and J. Wang, "A support vector machine with a hybrid kernel and minimal Vapnik-Chervonenkis dimension," *IEEE Transactions on Knowledge and Data Engineering,* vol. 16, pp. 385-395, APR 2004.

[29]    V. Vapnik, *The nature of statistical learning theory*. New York: Springer-Verlag, 1995.

[30]    K. R. Muller, S. Mika, G. Ratsch, K. Tsuda, and B. Scholkopf, "An introduction to kernel-based learning algorithms," *IEEE Transactions on Neural Networks,* vol. 12, pp. 181-201, MAR 2001.

[31]    A. Vinayagam, R. Konig, J. Moormann, F. Schubert, R. Eils, K. H. Glatting, and S. Suhai, "Applying support vector machines for gene ontology based gene function prediction," *Bmc Bioinformatics,* vol. 5, pp. -, AUG 26 2004.

[32]    S. Q. Le, T. B. Ho, and T. T. H. Phan, "A novel graph-based similarity measure for 2D chemical structures," *Genome Informatics,* vol. 14, pp. 82-91, 2004.

[33]    H. Kashima, K. Tsuda, and A. Inokuchi, "Marginalized kernels between labeled graphs," in *the 20th International Conference on Machine Learning*, 2003.

[34]    D. Haussler, "Convolution kernels on discrete structures," UC Santa Cruz1999.

[35]    B. J. Breitkreutz, C. Stark, and M. Tyers, "The GRID: The General Repository for Interaction Datasets," *Genome Biology,* vol. 4, 2003.

[36]    M. Ashburner, C. A. Ball, J. A. Blake, D. Botstein, H. Butler, J. M. Cherry, A. P. Davis, K. Dolinski, S. S. Dwight, J. T. Eppig, M. A. Harris, D. P. Hill, L. Issel-Tarver, A. Kasarskis, S. Lewis, J.

C. Matese, J. E. Richardson, M. Ringwald, G. M. Rubin, G. Sherlock, and G. O. Consortium, "Gene Ontology: tool for the unification of biology," *Nature Genetics,* vol. 25, pp. 25-29, MAY 2000.

[37] X. Li, H. Chen, Z. Huang, H. Su, and J. D. Martinez, "Global mapping of gene/protein interactions in PubMed abstracts: A framework and an experiment with P53 interactions," *J Biomed Inform,* Jan 17 2007.

[38] D. M. McDonald, H. Chen, H. Su, and B. B. Marshall, "Extracting gene pathway relations using a hybrid grammar: the Arizona Relation Parser," *Bioinformatics,* vol. 20., pp. 3370-3378, 2004.

[39] C.-C. Chang and C.-J. Lin, "LIBSVM: a library for support vector machines," in *http://www.csie.ntu.edu.tw/~cjlin/libsvm*, 2001.